\documentclass[aps,prb,twocolumn,floatfix,showpacs]{revtex4}
\usepackage{graphicx}
\begin{document}

%
\title{Planar and cagelike structures of gold clusters: Density-functional
pseudopotential calculations}

\author{Eva M. Fern\'{a}ndez}
\email[]{eva@lcb.fam.cie.uva.es} \affiliation{ Dpto.\ de  F\'\i sica
Te\'{o}rica, At\'{o}mica y \'{O}ptica, Universidad de Valladolid,
E-47011 Valladolid, Spain.}
\author{Jos\'{e} M. Soler}
\affiliation{Dpto.\ de F\'\i sica de la Materia Condensada, Universidad
Aut\'{o}noma de Madrid, E-28049 Madrid, Spain.}
\author{Luis C. Balbá\'{a}s}
\affiliation{Dpto.\ de F\'\i sica Te\'{o}rica, At\'{o}mica y \'{O}ptica,
Universidad de Valladolid, E-47011 Valladolid, Spain.}
\date{\today}

\begin{abstract}
We study why gold forms planar and cage-like clusters while copper and
silver do not. We use density functional theory and norm-conserving
pseudo-potentials with and without a scalar relativistic component. For
the exchange-correlation (xc) functional we use both the generalized
gradient (GGA) and the local density (LDA) approximations. We find that
planar Au$_n$ structures, with up to $n=11$, have lower energy than the
three-dimensional isomers only with scalar-relativistic pseudopotentials
and the GGA. In all other calculations, with more than 6 or 7 noble metal
atoms, we obtain three dimensional structures. However, as a general trend
we find that planar structures are more favorable with GGA than with LDA.
In the total energy balance, kinetic energy favors planar and cage
structures, while xc-energy favors 3D structures. As a second step, we
construct cluster structures having only surface atoms with O$_h$, T$_d$,
and I$_h$ symmetry. Then, assuming one valence electron per atom, we
select those with $2(l+1)^2$ electrons (with $l$ integer), which
correspond to the filling of a spherical electronic shell formed by
node-less one electron wave functions. Using scalar relativistic GGA
molecular dynamics at $T=600$K, we show that the cage-like structures of
neutral Au$_{32}$, Au$_{50}$, and Au$_{162}$ are meta-stable. Finally, we
calculate the static polarizability of the two lowest energy isomers of
Au$_n$ clusters as a means to discriminate isomers with planar (or
cage-like) geometry from those with compact structures. We also fit our
data to a semi-empirical relation for the size dependent polarizability
which involves the effective valence and the kinetic energy components for
the homogeneous and inhomogeneous electron density. Analyzing that fit, we
find that the dipole polarizability of gold clusters with planar and
cage-like structures corresponds to the linear response of 1.56
delocalized valence electrons, suggesting a strong screening of the
valence interactions due to the $d$-electrons.
\end{abstract}

\pacs{36.40.Cg, 36.40.Qv}

\maketitle

\section{\label{intro} Introduction}

Small clusters of metal atoms behave differently than the bulk matter,
because each additional atom, or even each additional electron, can
drastically change their electronic and geometrical properties
\cite{ekardt}. The noble metals clusters, with valence electron filling
$nd^{10}(n+1)s^1$, differ from the simple $s$-orbital alkali metals
\cite{pyykko04,pyykko05}, but also present striking differences among Cu,
Ag, and Au \cite{uzi-prl02,eva-prb04}. Well established structural
differences among Au$_n^{\nu}$ ($\nu=0,\pm1$) and Ag$_n^{\nu}$ or
Cu$_n^{\nu}$ clusters are the following: $i)$ Au$_n^{\nu}$ clusters,
specially the anions ($\nu=-1$), adopt planar structures up to larger
sizes than Ag and Cu clusters, as demonstrated by combined experimental
and theoretical studies~{\cite{furche,gilb,weis,yoon03}}; $ii)$
experimental photo-electron spectra for noble metal clusters with 55 atoms
\cite{moseler} indicate that silver and copper adopt some symmetry,
preferably icosahedral, whereas the pattern for Au$_{55}$ corresponds to
an amorphous structure~\cite{garzon1998}; $iii)$ anionic and neutral
Au$_{20}$ show a tetrahedral T$_d$ geometry \cite{Li03,wang03,eva-prb04},
but Ag$_{20}$ and Cu$_{20}$ have amorphous-like compact C$_s$ structures;
$iv)$ an icosahedral cage-like structure has been found to be very stable
for Au$_{32}$~\cite{johansson,prbAu32}, but not for silver and copper
{\cite{johansson,evaAu32}}. Other meta-stable cage-like structures for
gold clusters have been proposed recently \cite{evaAu32,jellinek05}.

The differences between Au and other noble metal clusters are usually
attributed to relativistic effects {\cite{uzi-prl02}}, which stabilize the
\emph{6s} orbital and destabilize the \emph{5d} one, favoring the
hybridization of these orbitals. However, although Pt shows as strong
relativistic effects as Au {\cite{pyykko04}}, it has been shown that the
competition between planar and 3D structures of Pt clusters is not
affected by relativistic effects {\cite{lixiao04}}. Notice that Pt$_7$ is
three-dimensional (3D) \cite{quan-tian04,quan-tian04b}, like Ag$_7$ or
Cu$_7$, but Au$_7$ is planar. Notice also that the largest $s$-orbital
contraction due to relativistic effects occurs in Au
\cite{schwerdtfeger02}. Consideration of the spin-orbit coupling does not
alter the relative stability of scalar relativistic structures of Au$_n$
clusters with $n \leq$ 20, but it increases the binding energy by about
0.08 eV/atom (1.85 kcal/mol) {\cite{xiao2004}}.

Comparison of density functional theory (DFT) results for Au$_6$ and
Au$_8$ at several levels of theory (that is, different
exchange-correlation functionals), with results from quantum chemical
calculations using second-order perturbation theory (MP2) or coupled
cluster methods (CCSD(T)), indicate that DFT predicts planar structures,
but MP2 and CCSD(T) predict the lowest energy Au$_8$ isomer to be
non-planar by 26.6 kcal/mol and 1.5 kcal/mol, respectively
{\cite{olson2005}}. Another recent calculation \cite{han06}, using
\emph{ab-initio} correlated-level theory, predicts Au$_8$ to be planar.

Concerning DFT calculations, we notice that the type of
exchange-correlation (xc) functional has a decisive influence on the
structural properties of gold clusters, but it is not so critical for
silver and copper clusters \cite{massobrio98}. Thus, first principles
calculations by means of the {\sc Siesta} code \cite{soler2002} with
scalar relativistic pseudopotentials, found planar structures of neutral
Au$_n$ clusters with up to 6 atoms using the local density approximation
(LDA) for the xc-functional \cite{soule-tetra04}, and up to 10 atoms using
the generalized gradient approximation (GGA) \cite{eva-prb04}. Such a
disimilar result was corroborated by a variety of different
pseudopotential and all-electron scalar relativistic DFT calculations
using LDA \cite{wang02,walker05} and/or GGA \cite{walker05,remacle05}.

In a recent paper, Gr$\ddot{\rm o}$nbeck and Broqvist \cite{gronbeck05}
compared the different contributions to the binding energy of several
planar and 3D structures of Au$_8$ and Cu$_8$ clusters, optimized within
GGA and LDA. They found that planar Au$_8$ isomers have a significative
smaller kinetic energy than 3D ones, which was attributed to $d$-electron
delocalization. A correlation between strong $s$-$d$ hybridization and
high stability of planar structures was found in ref. \onlinecite{uzi-prl02} for
noble metal heptamers, but does not appear to be a general tendency of
small Au clusters. Instead, the preference of planar configurations for
Au$_8$ isomers was attributed to a sizeable $d$-$d$ overlap and to
$d$-electron delocalization \cite{gronbeck05}.

Recently, the $\sigma$-aromaticity in saturated inorganic rings was
examined \cite{li05}. Evidence for $d$-orbital aromaticity in square
planar noble metal clusters \cite{wannere05} and in triangular gold rings
\cite{tsipis05} was presented also recently. The spherical
$\pi$-aromaticity of I$_h$ symmetrical gold fullerenes fulfilling a
generalized 2(\emph{l}+1)$^2$ $s$-electron rule \cite{hirsch00},
led Johansson \cite{johansson} to explain the
extra stability of a cage-like Au$_{32}$ cluster with I$_h$ symmetry
compared to other space-filling isomers.

A rough estimation of the photoabsorption response of several isomers of
Au$_{32}$ and Au$_{42}$ \cite{fa05b}, suggests that the cage-like
structures could be clearly distinguished from space filling isomers in
optical absorption experiments. However, it is difficult to separate in
these spectra the spectral features due to symmetry from the features due
to empty-cage effects. In this paper we will compare the calculated static
dipole polarizability of planar and cage-like Au$_n$ clusters with those
for 3D and compact isomers.

In section \ref{compu}, we outline the first principles method used in our
calculations. Results are presented and discussed in section
\ref{results}. In subsection \ref{planar}, we compare GGA and LDA
equilibrium structures of Au$_n$ clusters with 6 $\leq n \leq$ 9. We will
focus on the relation of 2D or 3D isomers with the delocalization of $d$
electrons, following the ideas of Gr$\ddot{\rm o}$nbeck and Broqvist
\cite{gronbeck05}. In subsection \ref{fuleoros} we contrast again LDA and
GGA predictions for the stability of {\it magic} cage-like structures
Au$_{18}$, Au$_{20}$, Au$_{32}$, Au$_{50}$, and Au$_{160}$, compared to
amorphous-like filling space isomers. The stability of these cage-like
structures against molecular dynamics at constant temperature (600 K), and
against the loss or gain of one electron is also tested. In subsection
\ref{pola} we investigate the use of the calculated static polarizability
as a physical property sensible to the cluster structure. In section
\ref{conclu} we will present our conclusions.

\section{\label{compu} Computational Procedure}

We use the first-principles code {\sc Siesta} \cite{soler2002} to solve
fully self-consistently the standard Kohn-Sham equations \cite{kohn-sham}
of DFT within the GGA as parametrized  by  Perdew, Burke and Ernzerhof
\cite{pbe96}, and within the LDA as parametrized  by Perdew and Zunger
\cite{pz81}. For each xc-approximation we use a norm conserving scalar
relativistic pseudopotential \cite{troullier91} in its fully nonlocal form
\cite{kleinman82}, generated from the Au atomic valence configurations
$5d^{10}6s^16p^0$, and core radii which we have tested and reported in
previous works \cite{eva-prb04,evaIJQC}. Flexible linear combinations of
numerical (pseudo) atomic orbitals are used as the basis set, allowing for
multiple-$\zeta$ and polarization orbitals. In order to limit the range of
the basis pseudoatomic orbitals (PAO), they are slightly excited by a
common energy shift (0.01 eV in this work), and truncated at the resulting
radial node \cite{sankey1989}. In the present calculations we used a
double-$\zeta$ $5p,6s$-basis, with maximum cutoff radius 7.62 Bohr. The
basis functions and the electron density are projected onto a uniform real
space grid in order to calculate the Hartree and exchange-correlation
potentials and matrix elements. The grid fineness is controlled by the
energy cutoff of the plane waves that can be represented in it without
aliasing (120 Ry in this work).

To obtain the equilibrium geometries, an unconstrained conjugate-gradient
structural relaxation using the DFT forces \cite{balbas01} was performed
for several initial cluster structures (typically more than ten),
suggested by the several geometries for Au$_n$, Au$_n^-$, and Au$_n^+$
isomers obtained previously \cite{eva-prb04}.

The static dipole polarizability of a cluster can be obtained by using the
standard numerical finite field perturbation method, in which the field
dependent energy is expanded with respect to an external uniform electric
field ${\bf F}$,
\begin{equation}
E = E^0 - \mu_iF_i - \frac{1}{2}\alpha_{ij}F_iF_j - \dots,
\end{equation}
where $i,j$ are cartesian coordinates and
the dipole moment and the static dipole polarizability are obtained as
energy derivatives,
 $\mu_i$ = $- \frac{\partial E}{\partial
F_i}|_{{\bf F}=0}$, and $\alpha_{ij}$ = $- \frac{\partial^2 E}{\partial
F_i \partial F_j}|_{{\bf F}=0}$, respectively. The external electric field
values used in our calculations were (in a.u.) $|{\bf F}|$ = 0.000, 0.001,
0.006, 0.010, 0.014, and 0.018. The energies calculated for these values
were fitted to a polynomial expansion to obtain the first-order and
second-order derivatives of energies with respect to the electric field
strength. The mean polarizability is calculated as, $\overline{\alpha} =
\mbox{Tr}(\alpha_{ij})/3$.

\section{\label{results} Results and Discussions}

\subsection{\label{planar} Planarity and $d$-electron delocalization
in Au$_n$}

\begin{figure}
\includegraphics[width=7cm]{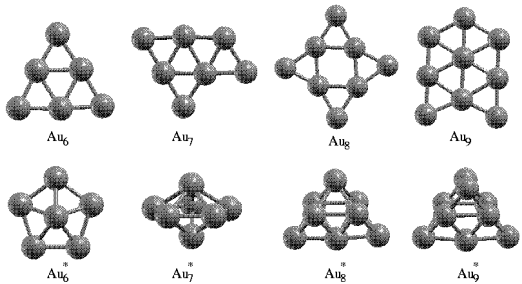}
\caption{\label{fig2a} (Color online) Equilibrium geometry of the lowest
energy isomers of gold clusters having planar (Au$_n$, upper row) or three
dimensional (Au$_n^*$, lower row) geometry as resulting from LDA and GGA
scalar relativistic calculations. For GGA the ground state is the planar
Au$_n$ isomer. For LDA the ground state is the 3D Au$_n^*$ isomer, except
for $n$ = 6, whose structure is also the planar Au$_6$.
}
\end{figure}

The onset of three-dimensional structures of neutral Au$_n$ clusters was
calculated at $n=11$ within GGA \cite{eva-prb04}, and at $n=7$ within LDA
\cite{soule-tetra04}. In this work we compare the planar and 3D lowest
energy isomers of Au$_n$ $(6 \leq n \leq 9)$ calculated within the {\sc
Siesta} code {\cite{soler2002}}, and using LDA \cite{pz81} and GGA
\cite{pbe96} for xc-functionals (with the corresponding LDA and GGA scalar
relativistic pseudo-potentials). Figure \ref{fig2a} shows our results for
the geometry of these isomers. The lower energy isomer is the planar one
for GGA, and the 3D one for LDA, except for $n$ = 6. Table \ref{table2}
gives the binding energy difference between 2D and 3D isomers, $\Delta E_b
= E_b(\mbox{Au}_n)-E_b(\mbox{Au}_n^*)$, for the relativistic and
non-relativistic calculations. We see that GGA leads to planar structures,
but LDA favors 3D structures for $n \geq$ 7 clusters. Thus, although the
planarity of Au$_7$ compared to the 3D structures of Ag$_7$ or Cu$_7$ was
attributed to relativistic effects \cite{uzi-prl02}, the observed
planarity of Au clusters \cite{furche,gilb,weis} is accounted for only
using the GGA theory. We have also optimized the Au$_n$ structures in
Figure \ref{fig2a} within GGA and LDA non relativistic pseudopotentials,
resulting that the 3D structures become more stable energetically than the
planar ones, except for $n$ = 6 within GGA, as shown in the last row of
Table \ref{table2}.

\begin{table}[h]
\caption{\label{table2} The binding energy difference $\Delta E_b =
E_b(\mbox{Au}_n)-E_b(\mbox{Au}_n^*)$, in eV, between the 2D and 3D isomers
of the gold clusters represented in Fig. \ref{fig2a}, optimized within LDA
and GGA xc-functionals using relativistic and non-relativistic (NR)
pseudopotentials.}
\begin{ruledtabular}
\begin{tabular}{lcccccccc}
\multicolumn{1}{c}{}&
\multicolumn{2}{c}{Au$_{6}$}&\multicolumn{2}{c}{Au$_{7}$}&
\multicolumn{2}{c}{Au$_{8}$} &\multicolumn{2}{c}{Au$_{9}$}\\
&LDA&GGA&LDA&GGA&LDA&GGA&LDA&GGA \\
\hline
$\Delta$E$_b$ & 0.14 & 0.15 & -0.02 & 0.03 & -0.06& 0.01 &-0.04& 0.01\\
$\Delta$E$_b^{NR}$ & -0.04 & 0.02 & -1.42 & -1.11 & -1.64& -1.37
&-1.29& -0.98
\end{tabular}
\end{ruledtabular}
\end{table}

Table \ref{table2a} gives the various energy differences (total, kinetic,
Coulomb, and exchange-correlation) between the second and first energy
isomers of scalar relativistic Au and Cu clusters with 6 $\leq n \leq$ 9
atoms. We will denote compact 3D structures with an asterisk, while its
absence indicates planar and cage-like geometries. The geometry of the Au
isomers is given in Figure \ref{fig2a}. The geometries of Cu$_6$ and
Cu$_6^*$ are similar to those of Au$_6$ and Au$_6^*$ respectively. For
Cu$_n$ with $n$ = 7, 8, 9, the geometries of the two lowest energy isomers
are both 3D, and are taken from our previous work \cite{eva-prb04}. We can
see in Table \ref{table2a} that planar structures have smaller kinetic
energy than 3D isomers, and larger for LDA than for GGA. Adding kinetic
and Coulomb energies, the 2D structures became more stable energetically
than the 3D ones. On the other hand, the xc-energy is more negative (it
contributes more to the binding energy) for 3D than for planar structures.
Both effects are stronger within the LDA but, in the balance of
total energy difference, the loss of kinetic energy in planar structures
dominates over the increase of xc-energy when using GGA, but the opposite
occurs with LDA. Thus, as a whole, GGA (LDA) favors 2D (3D) structures of
gold clusters. On the other hand, for Cu$_6$ and Cu$_6^*$ isomers, which
have the same geometry as Au$_6$ and Au$_6^*$, the change in kinetic and
Coulomb energies is not so noticeable as in gold.

For Cu$_n$ with $n$ = 7-9, whose first and second isomer geometries are
all 3D, the change in exchange-correlation energy, $\Delta_{xc}$, can be
positive or negative. For Cu$_8$, whose electronic properties can be
described approximately by the spherical jellium model \cite{eva-prb04},
we see that the sum of changes in kinetic and Coulomb energies roughly
cancel, resulting that the change in the total energy is ruled by the
change in xc-energy, as in the jellium model \cite{rubio93}.

The loss of kinetic energy in planar gold clusters with respect to their
3D isomers was attributed to electron delocalization \cite{gronbeck05},
but is not easy to reconcile that delocalization with the simultaneous
confinement in two dimensions. On the other hand, the xc energy becomes
less negative for planar configurations, which is also not clearly related
to delocalization and confinement in 2D gold clusters. We observe that the
calculated average bond length, $d_{av}$, is larger for 3D than for 2D
isomers. Specifically, the difference between $d_{av}$ of 3D and 2D Au$_n$
isomers with $n$ = 6, 7, 8, 9, is (in \AA ) 0.04, 0.11, 0.13, 0.09 (0.05,
0.11, 0.09, 0.09) for the LDA (GGA) calculation.

The kinetic energy of the electron gas in two dimensions is about one half
of the 3D case at the same density parameter, $r_s$, while the exchange
energy in 2D is slightly larger than in 3D \cite{lipparini03,glasser83}.
However, the correlation energy, at least in the RPA, is much larger in 2D
than in 3D \cite{lipparini03}. This consideration points again to the
importance of good exchange-correlation functionals in dealing with the
DFT structural description of gold clusters.

\begin{table}[h]
\caption{\label{table2a} Total, kinetic, Coulomb, and exchange-
correlation energy differences ($\Delta E_i = E_i(\mbox{Au}_n^*) -
E_i(\mbox{Au}_n)$), in eV, between the second isomer (3D, Au$_n^*$) and
first isomer (2D, Au$_n$) of gold clusters with the structures of Fig
\ref{fig2a}. For Cu$_n$ the first and second isomers are 3D, except for
Cu$_6$, where the two isomers are similar to Au$_6$ and Au$_6^*$. The
Cu$_n$ geometries are taken from ref. \onlinecite{eva-prb04}.
 }
\begin{ruledtabular}
\begin{tabular}{lrrrrrrrr}
 \multicolumn{1}{c}{}&
\multicolumn{2}{c}{$\Delta E_{tot}$}&\multicolumn{2}{c}{$\Delta E_{kin}$}&
\multicolumn{2}{c}{$\Delta E_{Coul}$} &\multicolumn{2}{c}{$\Delta E_{xc}$}\\
&LDA&GGA&LDA&GGA&LDA&GGA&LDA&GGA \\
\hline
 Au$_6$&0.86&0.88&11.01&10.63&-9.62&-9.42&-0.53&-0.33\\
Au$_7$&-0.15&0.24&12.49&10.34&-11.12&-9.52&-1.53&-0.57\\
Au$_8$&-0.46&0.04&7.37&7.03&-6.53&-6.80&-1.30&-0.19\\
Au$_9$&-0.34&0.07&5.07&4.63&-4.50&-4.52&-0.91&-0.04\\
\hline
Cu$_6$& &0.03& &1.03& &-078& &-0.22\\
Cu$_7$& &0.34& &0.20& &-0.98& &1.13\\
Cu$_8$& &0.89& &2.10& &-1.98& &0.77\\
Cu$_9$& &0.25& &-1.60& &1.87& &-0.02\\
\end{tabular}
\end{ruledtabular}
\end{table}

\subsection{\label{fuleoros}Magic cage-like structures of gold
clusters}

Stable cage-like structures of gold clusters have been predicted recently
for Au$_{32}$ \cite{johansson,prbAu32}, Au$_{26}$ \cite{fa05}, Au$_{42}$
{\cite{gao05}, and others \cite{evaAu32,jellinek05}. On the other hand,
although all atoms of Au$_{20}$ are at the surface, this cluster can be
considered as a small piece of bulk fcc gold \cite{Li03,wang03,eva-prb04}.

In this work we construct cage-like atomic structures starting with the
Platonic solids with triangular faces $-$ tetrahedron (T$_d$), octahedron
(O$_h$), and icosahedron (I$_h$) $-$ which are those allowing compact
planar packing. By adding atoms at the intersections of fcc planes in the
triangles, we obtain the following sequences for the number of atoms:
$n$(T$_d$)=4 + 2\emph{m}(\emph{m}+2), $n$(O$_h$)=6 +
4\emph{m}(\emph{m}+2), and $n$(I$_h$)=12 + 10\emph{m}(\emph{m}+2), where
\emph{m} = 0, 1, 2,$\dots$ is the number of atoms inserted in each edge.
When we add a central atom to each new triangle, we obtain cage-like
structures with $n$(T$_d$) = 2+6(\emph{m}+1)$^2$, $n$(O$_h$) =
2+12(\emph{m}+1)$^2$, and $n$(I$_h$) = 2+30(\emph{n}+1)$^2$. On the other
hand, from the electronic point of view, a cluster with nearly free
valence electrons is magic when it has filled electronic shells, having
well defined angular momentum, as in the jellium model. As we look for an
empty-cage with an approximately spherical surface, orbitals with radial
nodes have to be excluded and the only allowed electronic shells are
$1s,1p,1d,1f,\dots$ This leads to a magic number of electrons $n_e =
2(l+1)^2$, where $l$ is an integer number \cite{arom-rule}. Assuming that
each noble metal atom contributes with one valence electron, the equality
$n = 2(l+1)^2$ leads to the following magic neutral cage-like Au$_n$
clusters (containing less than 1000 atoms): $n$ = 8 (T$_d$), 18 (O$_h$),
32 (I$_h$), 50 (O$_h$), 98(T$_d$), 162 (I$_h$), and 578 (O$_h$). Double
anionic clusters should obey $n + 2 = 2(l+1)^2$ and they appear at $n=6$
and 198 (O$_h$) atoms. Double cationic magic clusters are the solutions of
$n - 2 = 2(l+1)^2$.
However, we exclude the T$_d$ clusters because they are far from
spherical. Also Au$_{20}$ can not be properly considered cage-like because
it contains many internal bonds, as similarly occurs to smaller clusters,
like Au$_{18}$. Nevertheless, in the following we will test the
estructural and electronic properties of Au$_n$ clusters with 18, 20, 32,
50, and 162 atoms.

\begin{figure}
\includegraphics[width=8cm]{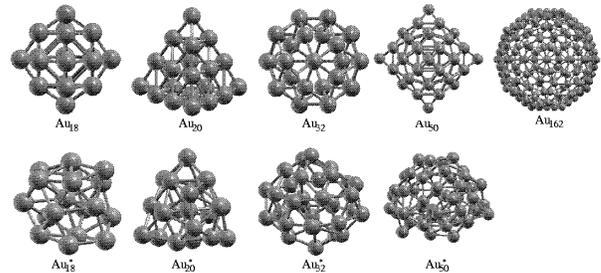}
\caption{(Color online) Cage-like (Au$_n$) and space filling (Au$_n^*$)
equilibrium isomeric structures of neutral gold clusters with $n$ = 18,
20, 32, 50, and 162, except Au$_{162}^*$ which is not optimized in this
paper. } \label{fig4}
\end{figure}

\begin{table}[h]
\caption{\label{table3} Binding energy difference between the cage-like
and compact structures of Fig \ref{fig4}, $\Delta E_b = E_b(\mbox{Au}_n) -
E_b(\mbox{Au}_n^*)$, optimized with the LDA and GGA approximations. }
\begin{ruledtabular}
\begin{tabular}{lcccccccc}
\multicolumn{1}{c}{}&\multicolumn{2}{c}{Au$_{18}$}&\multicolumn{2}{c}{Au$_{20}$}&
\multicolumn{2}{c}{Au$_{32}$} &\multicolumn{2}{c}{Au$_{50}$}
\\
 & LDA & GGA & LDA & GGA & LDA & GGA & LDA & GGA \\
\hline
$\Delta$E$_b$ & -0.07 & 0.01 & 0.01 & 0.03 & -0.06 & 0.01 & -0.14 & 0.01 \\
\end{tabular}
\end{ruledtabular}
\end{table}

We performed full relaxations of the initial cage-like magic structures
and several compact geometries obtained by forcing initially some surface
atoms inside these clusters. Fig \ref{fig4} shows the equilibrium
geometries with cage-like and compact structures, obtained after a non
exhaustive search, and optimized with forces $\leq$ 0.01 eV/\AA) at the
GGA and LDA levels. The cage-like equilibrium structures were proven to be
meta-stable after performing an ab-initio molecular dynamics run at
temperature of 600 K during 1000 steps, each of 2 fs. The binding energy
difference between cage-like and compact structures is tabulated in Table
\ref{table3}. We see that cage-like GGA structures are slightly more bound
than the compact ones. Instead, LDA leads to compact structures without
symmetry, except for Au$_{20}$. The true ground state is not known,
however, and improved functionals could lead to compact, ordered or
disordered, structures.

As a further test, we calculate the relative stability of cage-like and
compact isomers after loss or gain of one electron. Fig \ref{fig5} shows
the total energy difference per atom between cage-like and compact
equilibrium structures of cationic, neutral, and anionic clusters with
$n$=18, 20, 32 atoms. We see that only the cation Au$_{32}^+$ and the anion
Au$_{20}^-$ are still cage-like. Interestingly, the lowest energy isomer
of both ionic clusters, Au$_{18}^+$ and Au$_{18}^-$, is not
cage-like, contrary to the neutral Au$_{18}$.

In Fig \ref{fig6} we compare the density of states (DOS) of the cage-like
and amorphous structures of neutral Au$_n$ ($n$=18, 20). We see that the
cage-like geometries show a more structured DOS, with well defined peaks.
This is probably due to a higher geometrical symmetry (O$_h$ for Au$_{18}$
and T$_d$ for Au$_{20}$), as shown in a previous work for Au$_{55}$
\cite{moseler}. This fact is independent of the GGA or LDA level of
theory. The LDA DOS profile tends to be shifted (to lower energies for
Au$_{20}$) with respect to the GGA one, and the HOMO-LUMO gap is smaller
for LDA than for GGA. On the other hand the HOMO-LUMO gap is considerably
larger for cage-like Au$_{20}$ than for cage-like Au$_{18}$, indicating
that the former is much more stable than the later.

\begin{figure}[h]
\includegraphics[width=7cm]{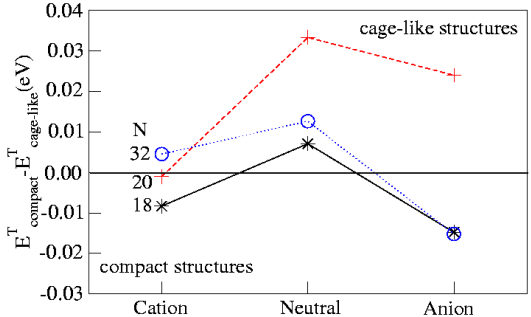}
\caption{(Color online) Total GGA energy difference per atom (in eV)
between the lower energy isomer of compact and cage-like structures for
cationic, neutral, and anionic clusters of gold with $n$=18 (stars), 20
(crosses), and 32 (circles) atoms.} \label{fig5}
\end{figure}

\begin{figure}
\includegraphics[width=7cm]{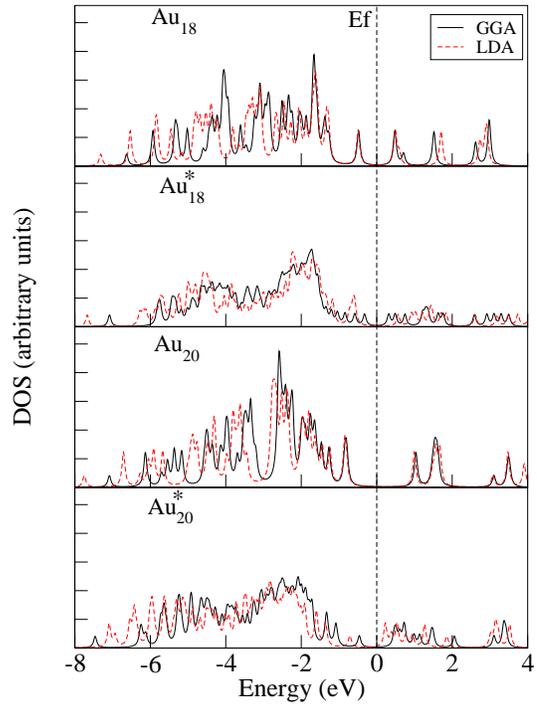}
\caption{(Color online) Density of states (DOS) of the lower energy
cage-like and amorphous isomers of Au clusters with 18 and 20 atoms,
calculated at the LDA and GGA levels of theory.}
 \label{fig6}
\end{figure}

\subsection{\label{pola} Static dipole polarizability}

The minimum polarizability principle states that any system evolves
naturally towards a state of minimum polarizability, but exceptions have
been reported \cite{torrent05}. In Table \ref{table5} are given the
results of our scalar-relativistic GGA calculations for the mean
polarizability per atom of the two lowest energy states of the Au clusters
reported in the subsections above.

The static dipole polarizability of atomic Au comes to 20.53 a.u.
(Bohr$^3$) for the scalar-relativistic calculation, and 33.06 a.u. for the
non-relativistic calculation, which is a clear manifestation of the
relativistic size contraction of gold \cite{pyykko04}. Our Au
polarizability is smaller than other calculated and experimental values
quoted in the literature \cite{pyykko04,roos05,neogrady97,castro04}.
Different estimates of the experimental value are 30$\pm$4 a.u.
\cite{pyykko04} and 39.1$\pm$9.8  a.u. \cite{roos05}. A quantum chemical
CCSD(T) calculation gives 36.06 a.u. \cite{neogrady97}, and a recent
CASSCF/CASPT2 relativistic calculation gives 27.9 a.u. \cite{roos05}.

For noble metal dimers, recent relativistic and non-relativistic
calculations \cite{saue03} found that the reduction of the polarizability,
due to the relativistic contraction effect, amounts to 39.8 \% for Au$_2$,
15.8 \% for Ag$_2$, and 6\% for Cu$_2$. From the TDDFT study of Castro and
coworkers \cite{castro04}, the static polarizability of Au$_n$ clusters up
to $n$ = 4 are affected up to $\sim$ 2\% by the inclusion of the
spin-orbit term.

From the TDDFT calculations of Castro \cite{castro04a}, the static
polarizability of the 3D tetrahedral isomer of Au$_4$ is 2.5\% higher than
the one for the planar D$_{2h}$ ground state. We obtain a larger
difference (15.5 \%) between our two planar Au$_4$ isomers (a rhombus and
a triangle with another Au on top, see ref \cite{eva-prb04}. Our result
for the Au$_n$ polarizability with \emph{n} = 2-9, is similar to the one
calculated by Zhao and coworkers \cite{zhao03} using a finite field
perturbation method like ours. However, the polarizability per atom
obtained by these authors for planar Au$_{18}$ and Au$_{20}$ is much
higher than ours for cage-like and compact isomers. On the other hand, the
difference between GGA and LDA polarizabilities of planar Au$_n$ clusters
($n$ = 2-20) calculated in ref. \onlinecite{zhao03} is less than 2\%.

\begin{table}[h]
\caption{\label{table5} The third and fourth columns give, respectively,
the mean polarizability per atom $\overline{\alpha}/n$ of the first and
second isomers of Au$_n$ clusters. The last column gives their difference.
For comparison, the first column gives the $\overline{\alpha}$/n value
calculated by Zhao and coworkers for planar estructures \cite{zhao03}. All
polarizabilities are in atomic units (bohr$^3$).}
\centering
\begin{ruledtabular}
\begin{tabular}{ccccr}
 &  & $\overline{\alpha}/n$ & $\overline{\alpha}/n$ &  \\
$n$ & Ref.~\onlinecite{zhao03} & 1st isomer & 2nd isomer &
$\Delta \overline{\alpha}~(\%)$ \\
\hline
2 & 29.4 & 28.3 & 29.0 & 2.5  \\
3 & 31.1 & 36.4 & 43.0 & 18.2  \\
4 & 32.7 & 31.5 & 37.2 & 17.9\\
5 & 34.2 & 32.9 & 46.6 & 41.6  \\
6 & 34.2 & 32.7 & 40.0 & 22.5  \\
7 & 34.3 & 33.9 & 30.9 & -9.0  \\
8 & 36.2 & 34.7 & 36.1 & 4.2 \\
9 & 38.5 & 37.8 & 48.1 & 27.3 \\
\hline
18& 42.7 & 32.1 & 31.4 & -2.4  \\
20& 43.7 & 35.2 & 32.9 & -6.4\\
32&  & 33.9 & 33.5  & -1.3\\
\end{tabular}
\end{ruledtabular}
\end{table}

We see in Table \ref{table5} that the average polarizability per atom of
Au$_n$ clusters is remarkably constant, with small odd-even effects for $n
\leq$ 7. Up to $n$ = 6 the first and second isomers are planar, and the
larger polarizability of the second isomer is due to the larger average
Au-Au bond length. The polarizability of the Au$^*_7$ 3D isomer is smaller
than that of the planar isomer, due to the compact pentagonal bipyramid
geometry, and it constitutes one of the exceptions to the minimum
polarizability principle \cite{torrent05}. The cage-like structures seem
to be another exception to that rule.

The average mean polarizability per atom for the first isomer of Au$_n$
clusters in Table \ref{table5} is $\overline{\alpha}_{av}$/$n$ = 33.57
a.u.. The jellium model for atomic clusters of monovalent metals fulfills
the relation $\alpha_{jel}$ = $r_s^3n$, where $r_s$ is the radius per
electron (in a.u.) of the bulk metal. For Au, with $r_s$ = 3.01 and the
GGA, we obtain the ratio $\tilde{\alpha} \equiv$
$\overline{\alpha}_{av}$/$\alpha_{jel}$ = 1.23. We can calculate a similar
ratio for Li, Na, Cu, and Ag clusters using results from experiments and
other calculations. Using the experimental values
\cite{benichou99,rayane99} for the dipole polarizability of Li ($r_s$ =
3.25), Na ($r_s$ = 3.93), and Cu ($r_s$ =2.67) clusters with $n \leq$ 10,
we obtain the ratios $\tilde{\alpha}_{exp}$(Li) = 2.26,
$\tilde{\alpha}_{exp}$(Na) = 2.11, and $\tilde{\alpha}_{exp}$(Li) = 2.0
for Li, Na, and Cu, respectively. From the GGA polarizability values
calculated in ref. \onlinecite{ghosh04} for Li and Na clusters with $n
\leq$ 10, we find $\tilde{\alpha}_{GGA}$(Li) = 2.47 and
$\tilde{\alpha}_{GGA}$(Na) = 1.81. For Cu clusters in the same range of
sizes, using the different GGA and LDA polarizability values reported by
Yang and Jackson \cite{yang05}, one has $\tilde{\alpha}_{GGA}$(Cu) =
1.93-1.98, and $\tilde{\alpha}_{LDA}$(Cu) = 1.79. Using recent
calculations \cite{jellinek05} for the polarizability of Ag clusters with
2 $\leq n \leq$ 8, results in $\tilde{\alpha}_{GGA}$(Ag) = 1.69. As a
whole, we see that the ratio $\overline{\alpha}_{av}$/$\alpha_{jel}$ is
considerably smaller for Au than for Li, Na, Cu, and Ag clusters.

The enhancement of dipole polarizabilty over the classical jellium model
(Mie value $\alpha_{jel}$) is directly proportional to the fraction of
electronic charge that extends beyond the positive background in the field
free system (spill-out)\cite{ekardt}. The smaller spill-out for gold with
respect to silver and copper can be attributed tentatively to the
relativistic contraction of the electronic cloud
\cite{pyykko04,schwerdtfeger02}.

We can explore a little more our calculated polarizability values using an
extended Thomas-Fermi-Weizs$\ddot{\rm a}$cker (TF$\lambda$W) jellium model
\cite{mananes94}, which predicts the mean polarizability of a cluster with
$n_e$ = $v$$n$ valence electrons ($v$ = valence, $n$ = number of atoms) as
\begin{equation}
\overline{\alpha} = \alpha_{jel}[1 +
3\frac{d(r_s)}{r_s}n_e^{-1/3}(1 +
\frac{\alpha_1(r_s)}{\alpha_0(r_s)}n_e^{-1/3})] \label{TFWpola}
\end{equation}
In this expression, $d(r_s)$ is the image plane position (the centroid of
the induced electron density for the flat metal surface), and $\alpha_1$
and $\alpha_0$ are two coefficients dependent of the parameter $\lambda$,
which take into account the weight of inhomogeneity-density correction
(Weizs$\ddot{\rm a}$cker term) to the Thomas-Fermi kinetic energy. In the
interval 3 $\leq r_s$ $\leq$ 4, it results $\alpha_1$($r_s$) $\simeq$ -0.1
a.u. independently of the $\lambda$ value (see Fig. 1 of ref.
\onlinecite{mananes94}), but $\alpha_0$($r_s$) is strongly dependent of
$\lambda$. In terms of a reduced polarizability, defined by $\alpha_{red}
= (\overline{\alpha}/\alpha_{jel} - 1)n^{2/3}$, we can write eq.
\ref{TFWpola} as
\begin{equation}
\alpha_{red} = A n^{1/3} + AB \label{polared}
\end{equation}
where $A = \frac{3}{v^{1/3}}\frac{d(r_s)}{r_s}$, and $B =
\frac{1}{v^{2/3}}\frac{\alpha_1(r_s)}{\alpha_0(r_s)}$. By fitting our data
in Table \ref{table5} to eqn. \ref{polared}, we obtain $\alpha_{red} =
1.15 n^{1/3} - 1.26$, and using $d$(Au)=1.34 a.u. for the image plane
distance of gold \cite{perdew88}, there results an effective valence $v$ =
1.56, and a ratio $\frac{\alpha_1(r_s)}{\alpha_0(r_s)}$ = -1.46. For
$\alpha_1(Au)$ $\simeq$ -0.1 we obtain $\alpha_0(Au)$ = 0.07.
Extrapolating the results in Fig. 1 of ref. \onlinecite{mananes94}, such a
small $\alpha_0(Au)$ value corresponds to an extremely small value of the
parameter $\lambda$, which means that the contribution of inhomogeneity
corrections to the kinetic energy is very small. This result agrees with
our conclusion in subsection \ref{planar} about the delocalization of
valence electrons in planar gold clusters. On the other hand, the
effective valence $v$ = 1.56 reflects in some way the screening of the
dipole response due to the \emph{d}-electrons \cite{castro04a}.

We test also an empirical linear relation between the cubic root of the
mean polarizability per atom, $\overline{\alpha}^{1/3}$/n, and the inverse
of the ionization potential per atom, I$_p^{-1}$/n. That relation was
tested in ref. \onlinecite{ghosh04} for Li and Na clusters with 2-10 atoms,
resulting for both cases (using calculated GGA values, in a.u.) in a
proportionality constant close to unity, with a linear correlation
coefficient 0.995. For the lower energy isomers of Au clusters reported in
Table \ref{table5}, there results a proportionality constant 0.925 and a
correlation coefficient of 0.984.

As probed by Yang and Jackson \cite{yang05}, temperature effects are a
possible source of discrepancy between calculated and measured
polarizabilities, because calculations are carried out for 0 K while
experiments are conducted at finite temperatures. The existence of a
permanent electric dipole in a cluster adds the following
temperature-dependent term to the effective polarizability
\cite{vasiliev97}:
\begin{equation}
\alpha_{eff} = \overline{\alpha} + \frac{\mu^2}{3kT}, \label{temp}
\end{equation}
where $\mu$ is the dipole moment and $k$ is the Boltzman constant. The
dipole contribution is important at low temperatures for clusters with
permanent dipole moments. Using the dipole moments calculated for our GGA
lower energy isomers, the correction of eqn. \ref{temp} to the mean
polarizability per atom is (in a.u) 1.39, 0.04, 0.74, 2.33, and 0.09, for
clusters with 3, 4, 5, 6, and 20 atoms, respectively. This small
correction still allows one to discriminate the polarizability of planar
and cage-like gold clusters from their isomers.

\section{\label{conclu} Conclusions}

We obtain that, using non-relativistic pseudopotentials, both GGA and LDA
predict the onset of three dimensional cluster structures already at $n$ =
6 for Cu$_n$ and Ag$_n$, and at $n$ = 7 (6) for Au$_n$. This result
changes by considering scalar-relativistic pseudopotentials within GGA,
resulting in planar Au$_n$ structures up to $n$ = 11.

From our scalar-relativistic results for the two lowest energy isomers of
Au$_n$ and Cu$_n$ with $n$ = 6-9 atoms, we find that planar structures
have smaller kinetic energy than 3D isomers, and this effect is much
larger for gold than for copper clusters. Adding kinetic and Coulomb
energies, the 2D structures became more stable than the 3D ones, and this
effect is more noticeable for LDA than for GGA. On the other hand, the
xc-energy is more negative (contribute more to the binding energy) for 3D
than for planar structures, and this effect is notably enhanced within
LDA. Thus, in the total energy balance, kinetic energy loss favors planar
GGA structures, but xc-energy favors LDA 3D structures.

 As a second step, we constructed clusters having only surface
atoms, and with O$_h$, T$_d$, and I$_h$ symmetry. From those, assuming one
valence electron per atom, we select the ones having 2(\emph{l}+1)$^2$
electrons, which correspond to the filling of a spherical electronic shell
formed by node-less one electron wave functions. We obtain, by means of
scalar-relativistic GGA calculations, that these cage-like structures for
neutral Au$_{18}$, Au$_{20}$, Au$_{32}$, Au$_{50}$, and Au$_{162}$ are
meta-stable after moderate (600 K) constant temperature molecular
dynamics. However, after addition or substraction of an electron, only the
anion Au$_{20}^{-}$ and the cation Au$_{32}^{+}$ remain cage-like.

Finally, we calculate the static polarizability of the two lowest energy
isomers of Au$_n$ clusters, which are planar ($n$ = 2-9) and cage-like
($n$ = 18, 20, 32) for the first isomer and planar ($n$ = 2-6) or space
filling 3D ($n$ = 7-9, 18, 20, 32) for the second isomer. In the range $n$
= 2-9, the polarizability per atom is smaller for the first isomer than
for the second, with the exception of $n$ = 7, confirming the empirical
rule of minimum polarizability. The contrary occurs for cage-like
structures, with larger $\bar{\alpha}$ than their space-filing isomers.

We fitted the polarizability of the first isomer of these gold clusters to
a semi-empirical relation between the cluster dipole polarizability and
its size, which involves the effective atomic valence and the kinetic
energy due to homogeneous and to inhomogeneous density components. From
that fit we extract a very small value for the kinetic energy component
due to inhomogeneous density, which suggest a delocalized character of the
valence electrons involved in the dipole response. We also obtain an
effective valence charge of 1.56 electrons, reflecting a dipole response
with strong screening of the $d$-electrons, as already reported previously
\cite{castro04a}.

\begin{acknowledgments}
We want to acknowledge the financial support from grants MAT2005-03415
and BFM2003-03372 of the Spanish Ministery of Science,
and from the FEDER of the European Community.
\end{acknowledgments}

\bibliography{fuleoro}

\begin{thebibliography}{64}
\expandafter\ifx\csname natexlab\endcsname\relax\def\natexlab#1{#1}\fi
\expandafter\ifx\csname bibnamefont\endcsname\relax
  \def\bibnamefont#1{#1}\fi
\expandafter\ifx\csname bibfnamefont\endcsname\relax
  \def\bibfnamefont#1{#1}\fi
\expandafter\ifx\csname citenamefont\endcsname\relax
  \def\citenamefont#1{#1}\fi
\expandafter\ifx\csname url\endcsname\relax
  \def\url#1{\texttt{#1}}\fi
\expandafter\ifx\csname urlprefix\endcsname\relax\def\urlprefix{URL }\fi
\providecommand{\bibinfo}[2]{#2}
\providecommand{\eprint}[2][]{\url{#2}}

\bibitem[{\citenamefont{Ekardt}()}]{ekardt}
\bibinfo{author}{\bibfnamefont{W.}~\bibnamefont{Ekardt}}, \eprint{\textit{Metal
  Clusters}, Wiley, Chichester, 1999}.

\bibitem[{\citenamefont{Pyykk$\ddot{\rm o}$}(2004)}]{pyykko04}
\bibinfo{author}{\bibfnamefont{P.}~\bibnamefont{Pyykk$\ddot{\rm o}$}},
  \bibinfo{journal}{Angew. Chem. Int. Ed.} \textbf{\bibinfo{volume}{43}},
  \bibinfo{pages}{4412} (\bibinfo{year}{2004}).

\bibitem[{\citenamefont{Pyykk$\ddot{\rm o}$}(2005)}]{pyykko05}
\bibinfo{author}{\bibfnamefont{P.}~\bibnamefont{Pyykk$\ddot{\rm o}$}},
  \bibinfo{journal}{Inorganica Chimica Acta} \textbf{\bibinfo{volume}{358}},
  \bibinfo{pages}{4113} (\bibinfo{year}{2005}).

\bibitem[{\citenamefont{Fern\'{a}ndez
  et~al.}(2004{\natexlab{a}})\citenamefont{Fern\'{a}ndez, Soler, Garz\'{o}n,
  and Balb\'{a}s}}]{eva-prb04}
\bibinfo{author}{\bibfnamefont{E.~M.} \bibnamefont{Fern\'{a}ndez}},
  \bibinfo{author}{\bibfnamefont{J.~M.} \bibnamefont{Soler}},
  \bibinfo{author}{\bibfnamefont{I.~L.} \bibnamefont{Garz\'{o}n}},
  \bibnamefont{and} \bibinfo{author}{\bibfnamefont{L.~C.}
  \bibnamefont{Balb\'{a}s}}, \bibinfo{journal}{Phys.\ Rev.\ B}
  \textbf{\bibinfo{volume}{70}}, \bibinfo{pages}{165403}
  (\bibinfo{year}{2004}{\natexlab{a}}).

\bibitem[{\citenamefont{H$\ddot{\rm a}$kkinen
  et~al.}(2002)\citenamefont{H$\ddot{\rm a}$kkinen, Moseler, and
  Landman}}]{uzi-prl02}
\bibinfo{author}{\bibfnamefont{H.}~\bibnamefont{H$\ddot{\rm a}$kkinen}},
  \bibinfo{author}{\bibfnamefont{M.}~\bibnamefont{Moseler}}, \bibnamefont{and}
  \bibinfo{author}{\bibfnamefont{U.}~\bibnamefont{Landman}},
  \bibinfo{journal}{Phys.\ Rev.\ Lett.} \textbf{\bibinfo{volume}{89}},
  \bibinfo{pages}{033401} (\bibinfo{year}{2002}).

\bibitem[{\citenamefont{Furche et~al.}(2002)\citenamefont{Furche, Ahlrich,
  Weis, Jacob, Gilb, Bienweiler, and Kappes}}]{furche}
\bibinfo{author}{\bibfnamefont{F.}~\bibnamefont{Furche}},
  \bibinfo{author}{\bibfnamefont{R.}~\bibnamefont{Ahlrich}},
  \bibinfo{author}{\bibfnamefont{P.}~\bibnamefont{Weis}},
  \bibinfo{author}{\bibfnamefont{C.}~\bibnamefont{Jacob}},
  \bibinfo{author}{\bibfnamefont{S.}~\bibnamefont{Gilb}},
  \bibinfo{author}{\bibfnamefont{T.}~\bibnamefont{Bienweiler}},
  \bibnamefont{and} \bibinfo{author}{\bibfnamefont{M.}~\bibnamefont{Kappes}},
  \bibinfo{journal}{J.\ Chem.\ Phys.} \textbf{\bibinfo{volume}{117}},
  \bibinfo{pages}{6982} (\bibinfo{year}{2002}).

\bibitem[{\citenamefont{S.~Gilb et~al.}(2002)\citenamefont{S.~Gilb, Furche,
  Ahlrichs, and Kappes}}]{gilb}
\bibinfo{author}{\bibfnamefont{P.~W.} \bibnamefont{S.~Gilb}},
  \bibinfo{author}{\bibfnamefont{F.}~\bibnamefont{Furche}},
  \bibinfo{author}{\bibfnamefont{R.}~\bibnamefont{Ahlrichs}}, \bibnamefont{and}
  \bibinfo{author}{\bibfnamefont{M.~M.} \bibnamefont{Kappes}},
  \bibinfo{journal}{J.\ Chem.\ Phys.} \textbf{\bibinfo{volume}{116}},
  \bibinfo{pages}{4094} (\bibinfo{year}{2002}).

\bibitem[{\citenamefont{Weis et~al.}(2002)\citenamefont{Weis, Bierweiler, Gilb,
  and Kappes}}]{weis}
\bibinfo{author}{\bibfnamefont{P.}~\bibnamefont{Weis}},
  \bibinfo{author}{\bibfnamefont{T.}~\bibnamefont{Bierweiler}},
  \bibinfo{author}{\bibfnamefont{S.}~\bibnamefont{Gilb}}, \bibnamefont{and}
  \bibinfo{author}{\bibfnamefont{M.~M.} \bibnamefont{Kappes}},
  \bibinfo{journal}{Chem.\ Phys.\ Lett.} \textbf{\bibinfo{volume}{355}},
  \bibinfo{pages}{355} (\bibinfo{year}{2002}).

\bibitem[{\citenamefont{H$\ddot{\rm a}$kkinen
  et~al.}(2003)\citenamefont{H$\ddot{\rm a}$kkinen, Yoon, Landman, Li, Zhai,
  and Wang}}]{yoon03}
\bibinfo{author}{\bibfnamefont{H.}~\bibnamefont{H$\ddot{\rm a}$kkinen}},
  \bibinfo{author}{\bibfnamefont{B.}~\bibnamefont{Yoon}},
  \bibinfo{author}{\bibfnamefont{U.}~\bibnamefont{Landman}},
  \bibinfo{author}{\bibfnamefont{X.}~\bibnamefont{Li}},
  \bibinfo{author}{\bibfnamefont{H.-J.} \bibnamefont{Zhai}}, \bibnamefont{and}
  \bibinfo{author}{\bibfnamefont{L.-S.} \bibnamefont{Wang}},
  \bibinfo{journal}{J. Phys. Chem. A} \textbf{\bibinfo{volume}{107}},
  \bibinfo{pages}{6168} (\bibinfo{year}{2003}).

\bibitem[{\citenamefont{H$\ddot{\rm a}$kkinen
  et~al.}(2004)\citenamefont{H$\ddot{\rm a}$kkinen, Moseler, Kostko, Morgner,
  Hoffmann, and von Issendorff}}]{moseler}
\bibinfo{author}{\bibfnamefont{H.}~\bibnamefont{H$\ddot{\rm a}$kkinen}},
  \bibinfo{author}{\bibfnamefont{M.}~\bibnamefont{Moseler}},
  \bibinfo{author}{\bibfnamefont{O.}~\bibnamefont{Kostko}},
  \bibinfo{author}{\bibfnamefont{N.}~\bibnamefont{Morgner}},
  \bibinfo{author}{\bibfnamefont{M.~A.} \bibnamefont{Hoffmann}},
  \bibnamefont{and} \bibinfo{author}{\bibfnamefont{B.}~\bibnamefont{von
  Issendorff}}, \bibinfo{journal}{Phys. Rev. Lett.}
  \textbf{\bibinfo{volume}{93}}, \bibinfo{pages}{093401}
  (\bibinfo{year}{2004}).

\bibitem[{\citenamefont{Garz\'on et~al.}(1998)\citenamefont{Garz\'on,
  Michaelian, Beltr\'an, Posada-Amarillas, Ordej\'on, Artacho, S\'achez-Portal,
  and Soler}}]{garzon1998}
\bibinfo{author}{\bibfnamefont{I.~L.} \bibnamefont{Garz\'on}},
  \bibinfo{author}{\bibfnamefont{K.}~\bibnamefont{Michaelian}},
  \bibinfo{author}{\bibfnamefont{M.~R.} \bibnamefont{Beltr\'an}},
  \bibinfo{author}{\bibfnamefont{A.}~\bibnamefont{Posada-Amarillas}},
  \bibinfo{author}{\bibfnamefont{P.}~\bibnamefont{Ordej\'on}},
  \bibinfo{author}{\bibfnamefont{E.}~\bibnamefont{Artacho}},
  \bibinfo{author}{\bibfnamefont{D.}~\bibnamefont{S\'achez-Portal}},
  \bibnamefont{and} \bibinfo{author}{\bibfnamefont{J.~M.} \bibnamefont{Soler}},
  \bibinfo{journal}{Phys. Rev. Lett.} \textbf{\bibinfo{volume}{81}},
  \bibinfo{pages}{1600} (\bibinfo{year}{1998}).

\bibitem[{\citenamefont{Li et~al.}(2003)\citenamefont{Li, Li, Zhai, and
  Wang}}]{Li03}
\bibinfo{author}{\bibfnamefont{J.}~\bibnamefont{Li}},
  \bibinfo{author}{\bibfnamefont{X.}~\bibnamefont{Li}},
  \bibinfo{author}{\bibfnamefont{H.-J.} \bibnamefont{Zhai}}, \bibnamefont{and}
  \bibinfo{author}{\bibfnamefont{L.-S.} \bibnamefont{Wang}},
  \bibinfo{journal}{Science} \textbf{\bibinfo{volume}{299}},
  \bibinfo{pages}{864} (\bibinfo{year}{2003}).

\bibitem[{\citenamefont{Wang et~al.}(2003)\citenamefont{Wang, Wang, and
  Zhao}}]{wang03}
\bibinfo{author}{\bibfnamefont{J.}~\bibnamefont{Wang}},
  \bibinfo{author}{\bibfnamefont{G.}~\bibnamefont{Wang}}, \bibnamefont{and}
  \bibinfo{author}{\bibfnamefont{J.}~\bibnamefont{Zhao}},
  \bibinfo{journal}{Chem. Phys. Lett.} \textbf{\bibinfo{volume}{380}},
  \bibinfo{pages}{716} (\bibinfo{year}{2003}).

\bibitem[{\citenamefont{Johansson et~al.}(2004)\citenamefont{Johansson,
  Sundholm, and Vaara}}]{johansson}
\bibinfo{author}{\bibfnamefont{M.~P.} \bibnamefont{Johansson}},
  \bibinfo{author}{\bibfnamefont{D.}~\bibnamefont{Sundholm}}, \bibnamefont{and}
  \bibinfo{author}{\bibfnamefont{J.}~\bibnamefont{Vaara}},
  \bibinfo{journal}{Angew. Chem. Int. Ed.} \textbf{\bibinfo{volume}{43}},
  \bibinfo{pages}{2678} (\bibinfo{year}{2004}).

\bibitem[{\citenamefont{X.~Gu et~al.}(2004)\citenamefont{X.~Gu, Wei, and
  Gong}}]{prbAu32}
\bibinfo{author}{\bibfnamefont{M.~J.} \bibnamefont{X.~Gu}},
  \bibinfo{author}{\bibfnamefont{S.~H.} \bibnamefont{Wei}}, \bibnamefont{and}
  \bibinfo{author}{\bibfnamefont{X.~G.} \bibnamefont{Gong}},
  \bibinfo{journal}{Phys. Rev. B} \textbf{\bibinfo{volume}{70}},
  \bibinfo{pages}{205401} (\bibinfo{year}{2004}).

\bibitem[{\citenamefont{Fern\'{a}ndez et~al.}(2006)\citenamefont{Fern\'{a}ndez,
  Torres, and Balb\'{a}s}}]{evaAu32}
\bibinfo{author}{\bibfnamefont{E.}~\bibnamefont{Fern\'{a}ndez}},
  \bibinfo{author}{\bibfnamefont{M.~B.} \bibnamefont{Torres}},
  \bibnamefont{and} \bibinfo{author}{\bibfnamefont{L.~C.}
  \bibnamefont{Balb\'{a}s}}, \bibinfo{journal}{Progr. Theor. Chem. Phys.
  (Springer)} \textbf{\bibinfo{volume}{15}}, \bibinfo{pages}{407}
  (\bibinfo{year}{2006}).

\bibitem[{\citenamefont{Wang et~al.}(2005)\citenamefont{Wang, Jellinek, Zhao,
  Chen, King, and v.~R.~Schleyer}}]{jellinek05}
\bibinfo{author}{\bibfnamefont{J.}~\bibnamefont{Wang}},
  \bibinfo{author}{\bibfnamefont{J.}~\bibnamefont{Jellinek}},
  \bibinfo{author}{\bibfnamefont{J.}~\bibnamefont{Zhao}},
  \bibinfo{author}{\bibfnamefont{Z.}~\bibnamefont{Chen}},
  \bibinfo{author}{\bibfnamefont{R.~B.} \bibnamefont{King}}, \bibnamefont{and}
  \bibinfo{author}{\bibfnamefont{P.}~\bibnamefont{v.~R.~Schleyer}},
  \bibinfo{journal}{J. Phys. Chem. A} \textbf{\bibinfo{volume}{109}},
  \bibinfo{pages}{9265} (\bibinfo{year}{2005}).

\bibitem[{\citenamefont{Xiao and Wang}(2004{\natexlab{a}})}]{lixiao04}
\bibinfo{author}{\bibfnamefont{L.}~\bibnamefont{Xiao}} \bibnamefont{and}
  \bibinfo{author}{\bibfnamefont{L.}~\bibnamefont{Wang}}, \bibinfo{journal}{J.
  Phys. Chem. A} \textbf{\bibinfo{volume}{108}}, \bibinfo{pages}{8605}
  (\bibinfo{year}{2004}{\natexlab{a}}).

\bibitem[{\citenamefont{Tian et~al.}(2004)\citenamefont{Tian, Ge, Sahu, Wang,
  Yamada, and Mashiko}}]{quan-tian04}
\bibinfo{author}{\bibfnamefont{W.~Q.} \bibnamefont{Tian}},
  \bibinfo{author}{\bibfnamefont{M.}~\bibnamefont{Ge}},
  \bibinfo{author}{\bibfnamefont{B.~R.} \bibnamefont{Sahu}},
  \bibinfo{author}{\bibfnamefont{D.}~\bibnamefont{Wang}},
  \bibinfo{author}{\bibfnamefont{T.}~\bibnamefont{Yamada}}, \bibnamefont{and}
  \bibinfo{author}{\bibfnamefont{S.}~\bibnamefont{Mashiko}},
  \bibinfo{journal}{J. Phys. Chem. A} \textbf{\bibinfo{volume}{108}},
  \bibinfo{pages}{3806} (\bibinfo{year}{2004}).

\bibitem[{\citenamefont{Tian et~al.}(2005)\citenamefont{Tian, Ge, Sahu, Wang,
  Yamada, and Mashiko}}]{quan-tian04b}
\bibinfo{author}{\bibfnamefont{W.~Q.} \bibnamefont{Tian}},
  \bibinfo{author}{\bibfnamefont{M.}~\bibnamefont{Ge}},
  \bibinfo{author}{\bibfnamefont{B.~R.} \bibnamefont{Sahu}},
  \bibinfo{author}{\bibfnamefont{D.}~\bibnamefont{Wang}},
  \bibinfo{author}{\bibfnamefont{T.}~\bibnamefont{Yamada}}, \bibnamefont{and}
  \bibinfo{author}{\bibfnamefont{S.}~\bibnamefont{Mashiko}},
  \bibinfo{journal}{J. Phys. Chem. A} \textbf{\bibinfo{volume}{109}},
  \bibinfo{pages}{6620} (\bibinfo{year}{2005}).

\bibitem[{\citenamefont{Schwerdtfeger}(2002)}]{schwerdtfeger02}
\bibinfo{author}{\bibfnamefont{P.}~\bibnamefont{Schwerdtfeger}},
  \bibinfo{journal}{Heteroatom Chemistry} \textbf{\bibinfo{volume}{13}}
  (\bibinfo{year}{2002}).

\bibitem[{\citenamefont{Xiao and Wang}(2004{\natexlab{b}})}]{xiao2004}
\bibinfo{author}{\bibfnamefont{L.}~\bibnamefont{Xiao}} \bibnamefont{and}
  \bibinfo{author}{\bibfnamefont{L.}~\bibnamefont{Wang}},
  \bibinfo{journal}{Chem. Phys. Lett.} \textbf{\bibinfo{volume}{392}},
  \bibinfo{pages}{452} (\bibinfo{year}{2004}{\natexlab{b}}).

\bibitem[{\citenamefont{Olson et~al.}(2005)\citenamefont{Olson, Varganov,
  Gordon, Metiu, Chretien, Piecuch, Kowalski, Kucharski, and
  Musial}}]{olson2005}
\bibinfo{author}{\bibfnamefont{R.~M.} \bibnamefont{Olson}},
  \bibinfo{author}{\bibfnamefont{S.}~\bibnamefont{Varganov}},
  \bibinfo{author}{\bibfnamefont{M.~S.} \bibnamefont{Gordon}},
  \bibinfo{author}{\bibfnamefont{H.}~\bibnamefont{Metiu}},
  \bibinfo{author}{\bibfnamefont{S.}~\bibnamefont{Chretien}},
  \bibinfo{author}{\bibfnamefont{P.}~\bibnamefont{Piecuch}},
  \bibinfo{author}{\bibfnamefont{K.}~\bibnamefont{Kowalski}},
  \bibinfo{author}{\bibfnamefont{S.}~\bibnamefont{Kucharski}},
  \bibnamefont{and} \bibinfo{author}{\bibfnamefont{M.}~\bibnamefont{Musial}},
  \bibinfo{journal}{J. Am. Chem. Soc.} \textbf{\bibinfo{volume}{127}},
  \bibinfo{pages}{1049} (\bibinfo{year}{2005}).

\bibitem[{\citenamefont{Han}(2006)}]{han06}
\bibinfo{author}{\bibfnamefont{V.~K.} \bibnamefont{Han}}, \bibinfo{journal}{J.
  Chem. Phys.} \textbf{\bibinfo{volume}{124}}, \bibinfo{pages}{024316}
  (\bibinfo{year}{2006}).

\bibitem[{\citenamefont{Massobrio et~al.}(1998)\citenamefont{Massobrio,
  Pasquarello, and Corso}}]{massobrio98}
\bibinfo{author}{\bibfnamefont{C.}~\bibnamefont{Massobrio}},
  \bibinfo{author}{\bibfnamefont{A.}~\bibnamefont{Pasquarello}},
  \bibnamefont{and} \bibinfo{author}{\bibfnamefont{A.~D.} \bibnamefont{Corso}},
  \bibinfo{journal}{J.\ Chem.\ Phys.} \textbf{\bibinfo{volume}{109}},
  \bibinfo{pages}{6626} (\bibinfo{year}{1998}).

\bibitem[{\citenamefont{Soler et~al.}(2002)\citenamefont{Soler, Artacho, Gale,
  Garc{\'\i}a, J.~Junquera, and S\'{a}nchez-Portal}}]{soler2002}
\bibinfo{author}{\bibfnamefont{J.~M.} \bibnamefont{Soler}},
  \bibinfo{author}{\bibfnamefont{E.}~\bibnamefont{Artacho}},
  \bibinfo{author}{\bibfnamefont{J.~D.} \bibnamefont{Gale}},
  \bibinfo{author}{\bibfnamefont{A.}~\bibnamefont{Garc{\'\i}a}},
  \bibinfo{author}{\bibfnamefont{P.~O.} \bibnamefont{J.~Junquera}},
  \bibnamefont{and}
  \bibinfo{author}{\bibfnamefont{D.}~\bibnamefont{S\'{a}nchez-Portal}},
  \bibinfo{journal}{J.\ Phys.:\ Condens.\ Matter}
  \textbf{\bibinfo{volume}{14}}, \bibinfo{pages}{2745} (\bibinfo{year}{2002}).

\bibitem[{\citenamefont{de~Bas et~al.}(2004)\citenamefont{de~Bas, Ford, and
  Cortie}}]{soule-tetra04}
\bibinfo{author}{\bibfnamefont{B.~S.} \bibnamefont{de~Bas}},
  \bibinfo{author}{\bibfnamefont{M.~J.} \bibnamefont{Ford}}, \bibnamefont{and}
  \bibinfo{author}{\bibfnamefont{M.~B.} \bibnamefont{Cortie}},
  \bibinfo{journal}{Journal of Molecular Structure (Theochem)}
  \textbf{\bibinfo{volume}{686}}, \bibinfo{pages}{193} (\bibinfo{year}{2004}).

\bibitem[{\citenamefont{Wang et~al.}(2002)\citenamefont{Wang, Wang, and
  Zhao}}]{wang02}
\bibinfo{author}{\bibfnamefont{J.}~\bibnamefont{Wang}},
  \bibinfo{author}{\bibfnamefont{G.}~\bibnamefont{Wang}}, \bibnamefont{and}
  \bibinfo{author}{\bibfnamefont{J.}~\bibnamefont{Zhao}},
  \bibinfo{journal}{Phys.\ Rev.\ B} \textbf{\bibinfo{volume}{66}},
  \bibinfo{pages}{035418} (\bibinfo{year}{2002}).

\bibitem[{\citenamefont{Walker}(2005)}]{walker05}
\bibinfo{author}{\bibfnamefont{A.~V.} \bibnamefont{Walker}},
  \bibinfo{journal}{J. Chem. Phys.} \textbf{\bibinfo{volume}{122}},
  \bibinfo{pages}{094310} (\bibinfo{year}{2005}).

\bibitem[{\citenamefont{Remacle and Kryachko}(2005)}]{remacle05}
\bibinfo{author}{\bibfnamefont{F.}~\bibnamefont{Remacle}} \bibnamefont{and}
  \bibinfo{author}{\bibfnamefont{E.~S.} \bibnamefont{Kryachko}},
  \bibinfo{journal}{J. Chem. Phys.} \textbf{\bibinfo{volume}{122}},
  \bibinfo{pages}{044304} (\bibinfo{year}{2005}).

\bibitem[{\citenamefont{Gr$\ddot{\rm o}$nbeck and Broqvist}(2005)}]{gronbeck05}
\bibinfo{author}{\bibfnamefont{H.}~\bibnamefont{Gr$\ddot{\rm o}$nbeck}}
  \bibnamefont{and} \bibinfo{author}{\bibfnamefont{P.}~\bibnamefont{Broqvist}},
  \bibinfo{journal}{Phys. Rev. B} \textbf{\bibinfo{volume}{71}},
  \bibinfo{pages}{73408} (\bibinfo{year}{2005}).

\bibitem[{\citenamefont{Li et~al.}(2005)\citenamefont{Li, Moran, Fan, and
  v.~R.~Schleyer}}]{li05}
\bibinfo{author}{\bibfnamefont{Z.-H.} \bibnamefont{Li}},
  \bibinfo{author}{\bibfnamefont{D.}~\bibnamefont{Moran}},
  \bibinfo{author}{\bibfnamefont{K.-N.} \bibnamefont{Fan}}, \bibnamefont{and}
  \bibinfo{author}{\bibfnamefont{P.}~\bibnamefont{v.~R.~Schleyer}},
  \bibinfo{journal}{J. Phys. Chem. A} \textbf{\bibinfo{volume}{109}},
  \bibinfo{pages}{3711} (\bibinfo{year}{2005}).

\bibitem[{\citenamefont{Wannere et~al.}(2005)\citenamefont{Wannere,
  Corminboeuf, Wang, Wodrich, King, and von R.~Schleyer}}]{wannere05}
\bibinfo{author}{\bibfnamefont{C.~S.} \bibnamefont{Wannere}},
  \bibinfo{author}{\bibfnamefont{C.}~\bibnamefont{Corminboeuf}},
  \bibinfo{author}{\bibfnamefont{Z.-X.} \bibnamefont{Wang}},
  \bibinfo{author}{\bibfnamefont{M.~D.} \bibnamefont{Wodrich}},
  \bibinfo{author}{\bibfnamefont{R.~B.} \bibnamefont{King}}, \bibnamefont{and}
  \bibinfo{author}{\bibfnamefont{P.}~\bibnamefont{von R.~Schleyer}},
  \bibinfo{journal}{J.\ Am.\ Chem.\ Soc.} \textbf{\bibinfo{volume}{127}},
  \bibinfo{pages}{5701} (\bibinfo{year}{2005}).

\bibitem[{\citenamefont{Tsipis and Tsipis}(2005)}]{tsipis05}
\bibinfo{author}{\bibfnamefont{A.~C.} \bibnamefont{Tsipis}} \bibnamefont{and}
  \bibinfo{author}{\bibfnamefont{C.~A.} \bibnamefont{Tsipis}},
  \bibinfo{journal}{J. Am. Chem. Soc.} \textbf{\bibinfo{volume}{127}},
  \bibinfo{pages}{10623} (\bibinfo{year}{2005}).

\bibitem[{\citenamefont{Hirsch et~al.}(2000)\citenamefont{Hirsch, Chen, and
  Jiao}}]{hirsch00}
\bibinfo{author}{\bibfnamefont{A.}~\bibnamefont{Hirsch}},
  \bibinfo{author}{\bibfnamefont{Z.}~\bibnamefont{Chen}}, \bibnamefont{and}
  \bibinfo{author}{\bibfnamefont{H.}~\bibnamefont{Jiao}},
  \bibinfo{journal}{Angew. Chem. Int. Ed.} \textbf{\bibinfo{volume}{39}},
  \bibinfo{pages}{3915} (\bibinfo{year}{2000}).

\bibitem[{\citenamefont{Fa et~al.}(2006)\citenamefont{Fa, Luo, and
  Dong}}]{fa05b}
\bibinfo{author}{\bibfnamefont{W.}~\bibnamefont{Fa}},
  \bibinfo{author}{\bibfnamefont{C.}~\bibnamefont{Luo}}, \bibnamefont{and}
  \bibinfo{author}{\bibfnamefont{J.}~\bibnamefont{Dong}},
  \bibinfo{journal}{Phys. Rev. B} \textbf{\bibinfo{volume}{73}},
  \bibinfo{pages}{085405} (\bibinfo{year}{2006}).

\bibitem[{\citenamefont{Kohn and Sham}(1965)}]{kohn-sham}
\bibinfo{author}{\bibfnamefont{W.}~\bibnamefont{Kohn}} \bibnamefont{and}
  \bibinfo{author}{\bibfnamefont{L.~J.} \bibnamefont{Sham}},
  \bibinfo{journal}{Phys.\ Rev.} \textbf{\bibinfo{volume}{145}},
  \bibinfo{pages}{561} (\bibinfo{year}{1965}).

\bibitem[{\citenamefont{Perdew et~al.}(1996)\citenamefont{Perdew, Burke, and
  Ernzerhof}}]{pbe96}
\bibinfo{author}{\bibfnamefont{J.~P.} \bibnamefont{Perdew}},
  \bibinfo{author}{\bibfnamefont{K.}~\bibnamefont{Burke}}, \bibnamefont{and}
  \bibinfo{author}{\bibfnamefont{M.}~\bibnamefont{Ernzerhof}},
  \bibinfo{journal}{Phys.\ Rev.\ Lett.} \textbf{\bibinfo{volume}{77}},
  \bibinfo{pages}{3865} (\bibinfo{year}{1996}).

\bibitem[{\citenamefont{Perdew and Zunger}(1981)}]{pz81}
\bibinfo{author}{\bibfnamefont{J.~P.} \bibnamefont{Perdew}} \bibnamefont{and}
  \bibinfo{author}{\bibfnamefont{A.}~\bibnamefont{Zunger}},
  \bibinfo{journal}{Phys.\ Rev.\ B} \textbf{\bibinfo{volume}{23}},
  \bibinfo{pages}{5075} (\bibinfo{year}{1981}).

\bibitem[{\citenamefont{Troullier and Mart´\'{i}ns}(1991)}]{troullier91}
\bibinfo{author}{\bibfnamefont{N.}~\bibnamefont{Troullier}} \bibnamefont{and}
  \bibinfo{author}{\bibfnamefont{J.~L.} \bibnamefont{Mart´\'{i}ns}},
  \bibinfo{journal}{Phys.\ Rev.\ B} \textbf{\bibinfo{volume}{43}},
  \bibinfo{pages}{1993} (\bibinfo{year}{1991}).

\bibitem[{\citenamefont{Kleinman and Bylander}(1982)}]{kleinman82}
\bibinfo{author}{\bibfnamefont{L.}~\bibnamefont{Kleinman}} \bibnamefont{and}
  \bibinfo{author}{\bibfnamefont{D.~M.} \bibnamefont{Bylander}},
  \bibinfo{journal}{Phys.\ Rev.\ Lett.} \textbf{\bibinfo{volume}{48}},
  \bibinfo{pages}{1425} (\bibinfo{year}{1982}).

\bibitem[{\citenamefont{Fern\'{a}ndez
  et~al.}(2004{\natexlab{b}})\citenamefont{Fern\'{a}ndez, Torres, and
  Balb\'{a}s}}]{evaIJQC}
\bibinfo{author}{\bibfnamefont{E.~M.} \bibnamefont{Fern\'{a}ndez}},
  \bibinfo{author}{\bibfnamefont{M.~B.} \bibnamefont{Torres}},
  \bibnamefont{and} \bibinfo{author}{\bibfnamefont{L.~C.}
  \bibnamefont{Balb\'{a}s}}, \bibinfo{journal}{Int. J. Quantum Chem.}
  \textbf{\bibinfo{volume}{99}}, \bibinfo{pages}{39}
  (\bibinfo{year}{2004}{\natexlab{b}}).

\bibitem[{\citenamefont{Sankey and Niklewski}(1989)}]{sankey1989}
\bibinfo{author}{\bibfnamefont{O.~F.} \bibnamefont{Sankey}} \bibnamefont{and}
  \bibinfo{author}{\bibfnamefont{D.~J.} \bibnamefont{Niklewski}},
  \bibinfo{journal}{Phys. Rev. B} \textbf{\bibinfo{volume}{40}},
  \bibinfo{pages}{3979} (\bibinfo{year}{1989}).

\bibitem[{\citenamefont{Balb\'{a}s et~al.}(2001)\citenamefont{Balb\'{a}s,
  Mart\'{\i}ns, and Soler}}]{balbas01}
\bibinfo{author}{\bibfnamefont{L.~C.} \bibnamefont{Balb\'{a}s}},
  \bibinfo{author}{\bibfnamefont{J.~L.} \bibnamefont{Mart\'{\i}ns}},
  \bibnamefont{and} \bibinfo{author}{\bibfnamefont{J.~M.} \bibnamefont{Soler}},
  \bibinfo{journal}{Phys.\ Rev.\ B} \textbf{\bibinfo{volume}{64}},
  \bibinfo{pages}{165110} (\bibinfo{year}{2001}).

\bibitem[{\citenamefont{Rubio et~al.}(1993)\citenamefont{Rubio, Balb\'as, and
  Alonso}}]{rubio93}
\bibinfo{author}{\bibfnamefont{A.}~\bibnamefont{Rubio}},
  \bibinfo{author}{\bibfnamefont{L.~C.} \bibnamefont{Balb\'as}},
  \bibnamefont{and} \bibinfo{author}{\bibfnamefont{J.~A.}
  \bibnamefont{Alonso}}, \bibinfo{journal}{Z. Physik D}
  \textbf{\bibinfo{volume}{26}}, \bibinfo{pages}{284} (\bibinfo{year}{1993}).

\bibitem[{\citenamefont{Lipparini}()}]{lipparini03}
\bibinfo{author}{\bibfnamefont{E.}~\bibnamefont{Lipparini}},
  \eprint{\textit{Modern Many-Particle Physics}, World Scientific, 2003, p.51}.

\bibitem[{\citenamefont{Glasser and Boersma}(1983)}]{glasser83}
\bibinfo{author}{\bibfnamefont{M.~L.} \bibnamefont{Glasser}} \bibnamefont{and}
  \bibinfo{author}{\bibfnamefont{J.}~\bibnamefont{Boersma}},
  \bibinfo{journal}{SIAM J. Appl. Math.} \textbf{\bibinfo{volume}{43}},
  \bibinfo{pages}{535} (\bibinfo{year}{1983}).

\bibitem[{\citenamefont{Fa et~al.}(2005)\citenamefont{Fa, Luo, and
  Dong}}]{fa05}
\bibinfo{author}{\bibfnamefont{W.}~\bibnamefont{Fa}},
  \bibinfo{author}{\bibfnamefont{C.}~\bibnamefont{Luo}}, \bibnamefont{and}
  \bibinfo{author}{\bibfnamefont{J.}~\bibnamefont{Dong}},
  \bibinfo{journal}{Phys. Rev. B} \textbf{\bibinfo{volume}{72}},
  \bibinfo{pages}{205428} (\bibinfo{year}{2005}).

\bibitem[{\citenamefont{Gao and Zeng}(2005)}]{gao05}
\bibinfo{author}{\bibfnamefont{Y.}~\bibnamefont{Gao}} \bibnamefont{and}
  \bibinfo{author}{\bibfnamefont{X.~C.} \bibnamefont{Zeng}},
  \bibinfo{journal}{J. Am. Chem. Soc.} \textbf{\bibinfo{volume}{127}},
  \bibinfo{pages}{3698} (\bibinfo{year}{2005}).

\bibitem[{aro()}]{arom-rule}
\eprint{Formally, this number of electrons is identical to the one allowed by
  the spherical aromaticity rule to constitute fullerens {\cite{hirsch00}}}.

\bibitem[{\citenamefont{M.Torrent-Sucarrat
  et~al.}(2005)\citenamefont{M.Torrent-Sucarrat, Duran, Luis, and
  Sol$\grave{\rm a}$}}]{torrent05}
\bibinfo{author}{\bibnamefont{M.Torrent-Sucarrat}},
  \bibinfo{author}{\bibfnamefont{M.}~\bibnamefont{Duran}},
  \bibinfo{author}{\bibfnamefont{J.~M.} \bibnamefont{Luis}}, \bibnamefont{and}
  \bibinfo{author}{\bibfnamefont{M.}~\bibnamefont{Sol$\grave{\rm a}$}},
  \bibinfo{journal}{J. Phys. Chem. A} \textbf{\bibinfo{volume}{109}},
  \bibinfo{pages}{615} (\bibinfo{year}{2005}).

\bibitem[{\citenamefont{Roos et~al.}(2005)\citenamefont{Roos, Lindh, Malmqvist,
  Veryazov, and Widmark}}]{roos05}
\bibinfo{author}{\bibfnamefont{B.~O.} \bibnamefont{Roos}},
  \bibinfo{author}{\bibfnamefont{R.}~\bibnamefont{Lindh}},
  \bibinfo{author}{\bibfnamefont{P.-A.} \bibnamefont{Malmqvist}},
  \bibinfo{author}{\bibfnamefont{V.}~\bibnamefont{Veryazov}}, \bibnamefont{and}
  \bibinfo{author}{\bibfnamefont{P.-O.} \bibnamefont{Widmark}},
  \bibinfo{journal}{J.\ Phys.\ Chem.} \textbf{\bibinfo{volume}{109}},
  \bibinfo{pages}{6575} (\bibinfo{year}{2005}).

\bibitem[{\citenamefont{Neogrady et~al.}(1997)\citenamefont{Neogrady, Kello,
  Urban, and Sadlej}}]{neogrady97}
\bibinfo{author}{\bibfnamefont{P.}~\bibnamefont{Neogrady}},
  \bibinfo{author}{\bibfnamefont{V.}~\bibnamefont{Kello}},
  \bibinfo{author}{\bibfnamefont{M.}~\bibnamefont{Urban}}, \bibnamefont{and}
  \bibinfo{author}{\bibfnamefont{A.~J.} \bibnamefont{Sadlej}},
  \bibinfo{journal}{Int. J. Quantum Chem.} \textbf{\bibinfo{volume}{63}},
  \bibinfo{pages}{557} (\bibinfo{year}{1997}).

\bibitem[{\citenamefont{Castro}(2004)}]{castro04}
\bibinfo{author}{\bibfnamefont{A.}~\bibnamefont{Castro}}, Ph.D. thesis,
  \bibinfo{school}{Universidad de Valladolid}, \bibinfo{address}{Spain}
  (\bibinfo{year}{2004}).

\bibitem[{\citenamefont{Saue and Jensen}(2003)}]{saue03}
\bibinfo{author}{\bibfnamefont{T.}~\bibnamefont{Saue}} \bibnamefont{and}
  \bibinfo{author}{\bibfnamefont{H.~J.~A.} \bibnamefont{Jensen}},
  \bibinfo{journal}{J.\ Chem.\ Phys.} \textbf{\bibinfo{volume}{118}},
  \bibinfo{pages}{522} (\bibinfo{year}{2003}).

\bibitem[{\citenamefont{Castro et~al.}(2004)\citenamefont{Castro, Marques,
  Alonso, and Rubio}}]{castro04a}
\bibinfo{author}{\bibfnamefont{A.}~\bibnamefont{Castro}},
  \bibinfo{author}{\bibfnamefont{M.~A.~L.} \bibnamefont{Marques}},
  \bibinfo{author}{\bibfnamefont{J.~A.} \bibnamefont{Alonso}},
  \bibnamefont{and} \bibinfo{author}{\bibfnamefont{A.}~\bibnamefont{Rubio}},
  \bibinfo{journal}{J. Comput. Nanosci.} \textbf{\bibinfo{volume}{1}},
  \bibinfo{pages}{231} (\bibinfo{year}{2004}).

\bibitem[{\citenamefont{Zhao et~al.}(2003)\citenamefont{Zhao, Yang, and
  Hou}}]{zhao03}
\bibinfo{author}{\bibfnamefont{J.}~\bibnamefont{Zhao}},
  \bibinfo{author}{\bibfnamefont{J.}~\bibnamefont{Yang}}, \bibnamefont{and}
  \bibinfo{author}{\bibfnamefont{J.~G.} \bibnamefont{Hou}},
  \bibinfo{journal}{Phys.\ Rev.\ B} \textbf{\bibinfo{volume}{67}},
  \bibinfo{pages}{085404} (\bibinfo{year}{2003}).

\bibitem[{\citenamefont{Benichou et~al.}(1999)\citenamefont{Benichou, Antoine,
  Rayane, Vezin, Dalby, Dugourd, Broyer, Ristori, Chandezon, Huber
  et~al.}}]{benichou99}
\bibinfo{author}{\bibfnamefont{E.}~\bibnamefont{Benichou}},
  \bibinfo{author}{\bibfnamefont{R.}~\bibnamefont{Antoine}},
  \bibinfo{author}{\bibfnamefont{D.}~\bibnamefont{Rayane}},
  \bibinfo{author}{\bibfnamefont{B.}~\bibnamefont{Vezin}},
  \bibinfo{author}{\bibfnamefont{F.~W.} \bibnamefont{Dalby}},
  \bibinfo{author}{\bibfnamefont{P.}~\bibnamefont{Dugourd}},
  \bibinfo{author}{\bibfnamefont{M.}~\bibnamefont{Broyer}},
  \bibinfo{author}{\bibfnamefont{C.}~\bibnamefont{Ristori}},
  \bibinfo{author}{\bibfnamefont{F.}~\bibnamefont{Chandezon}},
  \bibinfo{author}{\bibfnamefont{B.~A.} \bibnamefont{Huber}},
  \bibnamefont{et~al.}, \bibinfo{journal}{Phys.\ Rev.\ A}
  \textbf{\bibinfo{volume}{59}}, \bibinfo{pages}{R1} (\bibinfo{year}{1999}).

\bibitem[{\citenamefont{Rayane et~al.}(1999)\citenamefont{Rayane, Allouche,
  Benichou, Antoine, Aubert-Frecon, Dugourd, Broyer, Ristori, Chandezon,
  A.Huber et~al.}}]{rayane99}
\bibinfo{author}{\bibfnamefont{D.}~\bibnamefont{Rayane}},
  \bibinfo{author}{\bibfnamefont{A.~R.} \bibnamefont{Allouche}},
  \bibinfo{author}{\bibfnamefont{E.}~\bibnamefont{Benichou}},
  \bibinfo{author}{\bibfnamefont{R.}~\bibnamefont{Antoine}},
  \bibinfo{author}{\bibfnamefont{M.}~\bibnamefont{Aubert-Frecon}},
  \bibinfo{author}{\bibfnamefont{P.}~\bibnamefont{Dugourd}},
  \bibinfo{author}{\bibfnamefont{M.}~\bibnamefont{Broyer}},
  \bibinfo{author}{\bibfnamefont{C.}~\bibnamefont{Ristori}},
  \bibinfo{author}{\bibfnamefont{F.}~\bibnamefont{Chandezon}},
  \bibinfo{author}{\bibfnamefont{B.}~\bibnamefont{A.Huber}},
  \bibnamefont{et~al.}, \bibinfo{journal}{Eur. Phys. J. D}
  \textbf{\bibinfo{volume}{9}}, \bibinfo{pages}{243} (\bibinfo{year}{1999}).

\bibitem[{\citenamefont{Chandrakumar et~al.}(2004)\citenamefont{Chandrakumar,
  Ghanty, and Ghosh}}]{ghosh04}
\bibinfo{author}{\bibfnamefont{K.~R.~S.} \bibnamefont{Chandrakumar}},
  \bibinfo{author}{\bibfnamefont{T.~K.} \bibnamefont{Ghanty}},
  \bibnamefont{and} \bibinfo{author}{\bibfnamefont{S.~K.} \bibnamefont{Ghosh}},
  \bibinfo{journal}{J. Phys. Chem. A} \textbf{\bibinfo{volume}{108}},
  \bibinfo{pages}{6661} (\bibinfo{year}{2004}).

\bibitem[{\citenamefont{Yang and Jackson}(2005)}]{yang05}
\bibinfo{author}{\bibfnamefont{M.}~\bibnamefont{Yang}} \bibnamefont{and}
  \bibinfo{author}{\bibfnamefont{K.~A.} \bibnamefont{Jackson}},
  \bibinfo{journal}{J.\ Chem.\ Phys.} \textbf{\bibinfo{volume}{122}},
  \bibinfo{pages}{184317} (\bibinfo{year}{2005}).

\bibitem[{\citenamefont{Ma$\tilde{\rm n}$anes
  et~al.}(1994)\citenamefont{Ma$\tilde{\rm n}$anes, Membrado, Pacheco,
  Sa$\tilde{\rm n}$udo, and Balb\'{a}s}}]{mananes94}
\bibinfo{author}{\bibfnamefont{A.}~\bibnamefont{Ma$\tilde{\rm n}$anes}},
  \bibinfo{author}{\bibfnamefont{M.}~\bibnamefont{Membrado}},
  \bibinfo{author}{\bibfnamefont{A.~F.} \bibnamefont{Pacheco}},
  \bibinfo{author}{\bibfnamefont{J.}~\bibnamefont{Sa$\tilde{\rm n}$udo}},
  \bibnamefont{and} \bibinfo{author}{\bibfnamefont{L.~C.}
  \bibnamefont{Balb\'{a}s}}, \bibinfo{journal}{Int. J. Quantum Chem.}
  \textbf{\bibinfo{volume}{52}}, \bibinfo{pages}{767} (\bibinfo{year}{1994}).

\bibitem[{\citenamefont{Perdew}(1988)}]{perdew88}
\bibinfo{author}{\bibfnamefont{J.~P.} \bibnamefont{Perdew}},
  \bibinfo{journal}{Phys.\ Rev.\ B} \textbf{\bibinfo{volume}{37}},
  \bibinfo{pages}{6175} (\bibinfo{year}{1988}).

\bibitem[{\citenamefont{Vasiliev et~al.}(1997)\citenamefont{Vasiliev,
  $\ddot{\rm O}\breve{\rm g}\ddot{\rm u}$t, and Chelikowsky}}]{vasiliev97}
\bibinfo{author}{\bibfnamefont{I.}~\bibnamefont{Vasiliev}},
  \bibinfo{author}{\bibfnamefont{S.}~\bibnamefont{$\ddot{\rm O}\breve{\rm
  g}\ddot{\rm u}$t}}, \bibnamefont{and} \bibinfo{author}{\bibfnamefont{J.~R.}
  \bibnamefont{Chelikowsky}}, \bibinfo{journal}{Phys.\ Rev.\ Lett.}
  \textbf{\bibinfo{volume}{78}}, \bibinfo{pages}{4805} (\bibinfo{year}{1997}).

\end{thebibliography}

\end{document}